# Ultralow-$V_\pi L$ Silicon Electro-Optic Directional Coupler Switch with a Liquid Crystal Cladding

Li-Yuan Chiang, Chun-Ta Wang, Steve Pappert, and Paul K. L. Yu

*Abstract*—An ultralow-$V_\pi L$ photonic switch device is demonstrated utilizing the high optical and electrical field confinement in silicon slot waveguides coupled with the strong electro-optic response of nematic liquid crystals. A silicon photonic directional coupler switch with a modulation efficiency of 0.0195 V·mm and a loss-efficiency product of 0.0624 V·dB is achieved. The 1.5 mm long device is based on two-mode interference within a single slot waveguide resulting in a $V_\pi$ of 0.013 V and an extinction ratio of ~ 9dB at 1550 nm wavelength. The power consumption of the photonic switch is estimated to be below 0.6 nW and it possesses a response time of <1.5 ms. A comparative performance study between the directional coupler switch and a Mach-Zehnder Interferometer (MZI) switch is performed. The directional coupler switch is projected to have a smaller footprint and lower optical loss compared to a similar design MZI switch making it a strong candidate for switch matrix designs and applications.

*Index Terms*—Silicon photonics, optical switches, optical beam steering, slot waveguides, silicon-on-insulator, liquid crystals

## I. INTRODUCTION

SILICON photonic switches [1] have recently shown great potential to form large-scale switch matrices for optical signal processing applications including chip-scale Light Detection and Ranging (LIDAR) [2]-[4]. Consisting of an optical switch matrix, an array of emitters, and a collimating lens, chip-scale optical beam steering can be realized with lower power consumption and simplified electronic control compared with optical phased arrays [5]. For this switch-based beamforming approach, modulation efficiency ($V_\pi L$), footprint and optical loss are the most critical criteria for optical switches due to the cascaded fabric and large number of elements [6]. A miniaturized footprint enables a larger port count, reduced fabrication cost, and lower propagation loss. An ultralow-$V_\pi L$ photonic switch element can support sub-volt operation with short device length, facilitating compact, cost-effective, low-loss, and power-efficient switch matrices and optical beamformers.

Silicon-organic hybrid devices have made remarkable progress in achieving improved modulation efficiency (lower $V_\pi L$ product). Among the reported silicon-organic devices, Electro-Optic (EO) polymers were utilized to demonstrate high-speed and efficient modulators for optical communication applications [7]-[10], whereas nematic liquid crystals (NLCs) provide even stronger but slower EO responses [11]-[13]. By incorporating an NLC cladding with a silicon conductive slot waveguide Mach-Zehnder interferometer (MZI), a $V_\pi L$ as low as 0.0224 V·mm was demonstrated [14].

Plasmonic-organic hybrid (POH) devices are another way to realize ultralow $V_\pi L$. Combining plasmonic slot waveguides with highly efficient EO polymers, a $V_\pi L$ of 0.05 V·mm was demonstrated [15]. However, the high optical loss of the POH structures results in loss-efficiency products ($\alpha V_\pi L$, where $\alpha$ is the optical propagation loss of the device region in dB/mm) larger than 20 V·dB [15]. In comparison, the NLC-cladded silicon slot waveguide MZI switch was demonstrated with a $\alpha V_\pi L$ of ~0.25 V·dB [14].

In this paper, we propose and demonstrate a silicon photonic switch based on an NLC-cladded slot waveguide directional coupler (DC). We show a $V_\pi L$ of 0.0195 V·mm at an operating point of 0.935 V root-mean-square voltage for wavelengths near 1550 nm. This device offers the potential of sub-volt $V_\pi$ switching operation with a device length as short as 20 μm. We compare the DC switch configuration with the MZI switch configuration and display the DC competitiveness for designing compact and low-loss switch elements.

## II. DEVICE DESIGN AND OPERATION PRINCIPLE

In Fig. 1 we show the proposed silicon photonic switch based on a slot waveguide DC with an NLC cladding filling the slot region [16]. A strip-loaded conductive silicon slot waveguide structure [17] enables a highly confined external electric field

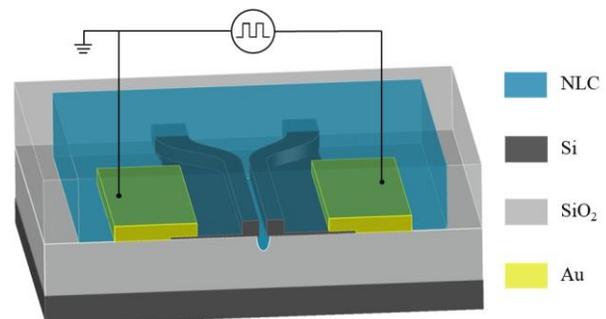

Fig. 1. Schematic of the proposed silicon EO switch. The slot waveguide directional coupler is clad with a nematic liquid crystal (NLC) layer.

This project has been funded by the Office of Naval Research under Contract # N00014-18-12027.

L.-Y. Chiang, S. Pappert, and Paul K. L. Yu are with the Department of Electrical and Computer Engineering, University of California San Diego, La Jolla, CA 92122, USA (e-mail: l1chiang@eng.ucsd.edu; spappert@eng.ucsd.edu; pyu@ucsd.edu).

C.-T. Wang is with the Department of Photonics, National Sun Yat Sen University, Kaohsiung 80424, Taiwan (e-mail: ctwang@mail.nsysu.edu.tw).

across the slot, allowing efficient refractive index tuning of the NLC in the slot region through reorienting the rod-shape molecules. The NLC material employed in this work is E7 (made by Merck), which has a large birefringence ($\Delta n \sim 0.2$) contributing to the strong EO response. Two optical modes, TE0 and TE1, propagating in the slot waveguide DC are launched by a single-mode TE-polarized light at a wavelength near 1550 nm from either one of the input waveguides. The TE0 and TE1 mode profiles as shown in Fig. 2 have distinct mode overlaps at the slot region, leading to the contrasting effective index changes while tuning the slot material index. Optical switching is achieved by tuning the two-mode interference condition,

$$L_{cc} = \frac{\lambda}{(n_0 - n_1)}, \quad (1)$$

where $\lambda$ is the light wavelength, $L_{cc}$ is the cross-coupling length, $n_0$ and $n_1$ are the electric-field dependent effective indices of the TE0 and TE1 modes, respectively. The waveguide geometry dependent switching performance is systematically studied using Finite Difference Eigenmode (FDE) simulation and summarily reported in [18]. Based on FDE simulation results, the waveguide geometry for device fabrication was selected.

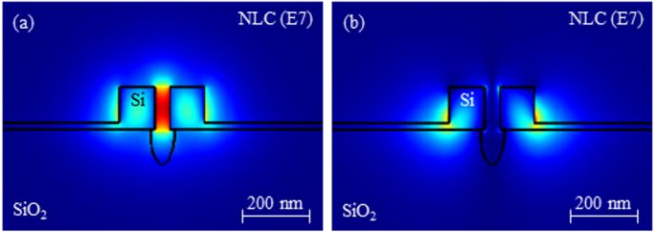

Fig. 2. Optical mode profiles of the (a) TE0 mode and (b) TE1 mode propagating in the cross-coupling slot waveguide

### III. DEVICE FABRICATION

The photonic switch was fabricated using a silicon-on-insulator (SOI) wafer with a 250-nm silicon layer on top of a 3-μm buried oxide layer. The silicon slot waveguide structures were patterned using electron-beam lithography and formed using reactive ion etching with SF6:CHF3 plasma. The slot waveguide has a 1.5-mm long and 115-nm wide slot region. The waveguide width is 265 nm for both silicon coupled arms. The silicon slabs have a thickness of 45 nm. The silicon coupled arms and slabs were n-doped at $10^{17}/cm^3$. The waveguide structures are clad with a 2-μm SiO2 layer deposited by plasma-enhanced chemical vapor deposition (PECVD). The slot region was opened and overetched through a combination of RIE and diluted buffered oxide etch (BOE). Gold contacts were formed by sputtering and lift-off processes. Fig. 3(a) and Fig. 3(b) show the resulting top view and slot cross section of the structure, respectively, under scanning electron microscope (SEM). E7 NLC was deposited on the chip as the final step to fill the slot and overetched region, utilizing its fluidic nature.

The slot region was intentionally overetched during fabrication. The overetch ensures the slot was cleared from remaining PECVD SiO2 and ready for NLC filling. Also, since the solid boundaries anchor a portion of NLC molecules and make them hard to be reoriented [19], keeping the bottom SiO2 boundary away from the bottom of the slot allows better control of the NLC molecules throughout the Si slot.

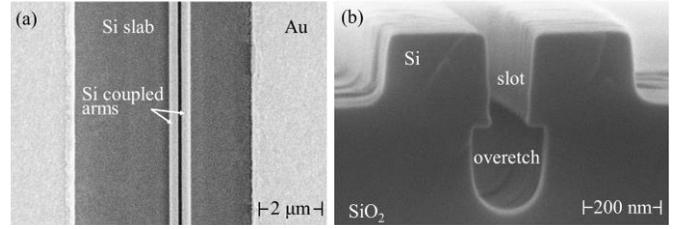

Fig. 3. SEM images of the proposed device displaying (a) top view and (b) cross section before applying an NLC cladding.

### IV. DEVICE CHARACTERIZATION

The photonic switch was driven with 5 kHz square-wave alternating current (ac) voltage to avoid degradation of the NLC material in the slot region [20]. The molecular directions of the NLC in the slot were tuned by changing the amplitude of the square-wave signal to achieve optical switching. The optical transmission and switching results as a function of the applied root-mean-square voltage measured from one of the output ports are shown in Fig. 4. Operating at 0.935 V, a $V_\pi$ of 0.013 V was measured as shown in Fig. 4(a). The corresponding $V_\pi L$ is 0.0195 V·mm. The device was switched on and off 10 cycles with the 0.928 V and 0.941 V applied as shown in Fig. 4(b). The device was also measured with 100 switching cycles. The standard deviation values calculated from the 100 on-state and off-state signal levels are 0.17 dB and 0.11 dB, respectively. The extinction ratio measured is ~9 dB. The temporal response of on/off switching between 0.928 V and 0.941 V measured from the device and the probe tips, respectively, are shown in Fig. 5. The measured 10%-90% rise time is 1.41 ms, and the 90%-10% fall time is 1.37 ms for the device, both limited by the electrical driving. The on-chip loss of the DC switch is -4.8 ± 0.8 dB. The $aV_\pi L$ calculated is therefore 0.0624 V·dB. We expect better extinction ratio and optical loss can be achieved with optimization of fabrication processes and coupling between the input/output waveguides and the slot waveguide.

### V. DISCUSSION

In addition to a DC having TE0 and TE1 mode interference, the slot waveguide configuration can also be configured into an MZI having interference of two TE0 modes in separate slot waveguides. The demonstrated performance of our DC switch was compared to that of a reported slot waveguide MZI switch [14] and listed in Table 1. Using the same NLC material E7, the DC switch achieved smaller $V_\pi L$ and on-chip loss. Regarding the operating point and switching time of the DC and MZI devices, since they are limited by the NLC material properties, there is no significant difference.

The improved efficiency of the DC switch originates from the optimization of slot waveguide geometry, instead of the physical nature of the device architecture. The modulation efficiency is highly dependent on both the slot waveguide geometry and NLC material properties. From our previous simulation results [18], increased Si waveguide height (up to

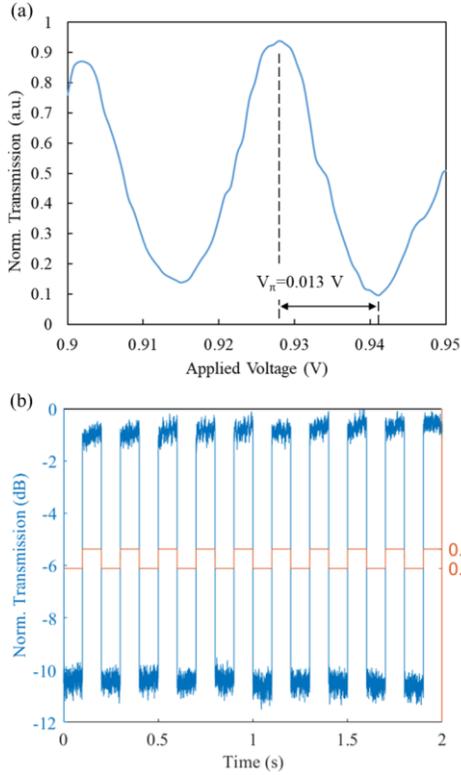

Fig. 4. (a) Normalized optical transmission of the device as a function of the applied voltage (b) Real-time trace of the device's optical transmission during 10 switching on and off cycles shown with the corresponding applied voltage.

330 nm), narrower slot width, and lower slab height provide larger mode overlap of the TE0 mode in the slot region. With larger mode overlap confined in the slot region, the effective index of the TE0 mode is more sensitive to slot index tuning. The effect of material properties on device performance is more complex. Using the same NLC material, variations of the anchoring strength, sidewall roughness, and slot geometry can affect the EO modulation efficiency [21], [22].

In principle, with the same slot waveguide geometry and NLC material properties, an MZI switch has a higher efficiency compared with that of a DC switch [18]. This is due to the unwanted and inevitable modulation in the TE1 mode of DC switches. Despite having better efficiency, the MZI is not always a better option to configure photonic switches. Combiner/splitter, additional routing waveguides, and strip-to-slot mode convertors are required at both the input and output side of an MZI switch, contributing excess optical loss and device footprint. These elements are not scalable with improved efficiency and shorter slot waveguide length. For ultracompact switch design, the DC can be preferable due to its potentially smaller footprint and lower optical loss. For example, based on simulations and the demonstrated $V_\pi L$ of 0.0195 V·mm from the DC switch, we expect a $V_\pi L$ of 0.0142 V·mm can be achieved if the device is patterned as an MZI switch. The values correspond to 19.5 μm and 14.2 μm of device lengths for the DC and MZI, respectively, operated with 1 V $V_\pi$. However, this is not the complete story. Whereas 5.3 μm shorter in the estimated slot waveguide region, typical MZI switches include a pair of ~30 μm long multi-mode interferometers (MMIs) [23], ~5 μm long additional routing waveguides, and ~5 μm long strip-to-slot converters [24], making it extremely challenging to design an MZI switch with a total length in the tens of micrometers. Therefore, for switches requiring small footprint, the advantage of better efficiency in an MZI device is no longer the dominant design consideration.

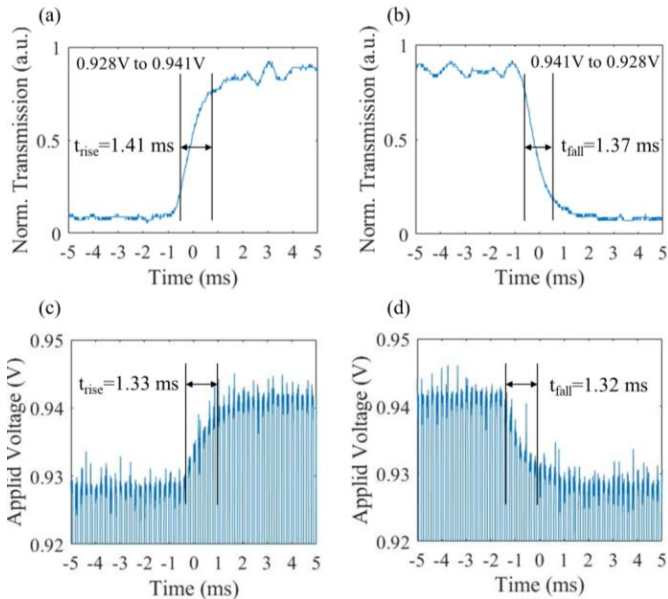

Fig. 5. Measured temporal responses of (a) switching on and (b) switching off from the device. Electrical driving of (c) switching on and (d) switching off measured from the probe tips.

TABLE I
COMPARISON OF DEMONSTRATED NLC-FILLED SLOT WAVEGUIDE SWITCHES

| Device Architecture | Slot WG DC switch (this work) | Slot WG MZI switch [14] |
|---|---|---|
| NLC material | E7 | E7 |
| $V_\pi L$ (V·mm) | 0.0195 | 0.0224 |
| Operating Point (V) | 0.935 | ~1.0 |
| $V_\pi$ (V) | 0.013 | 0.0224 |
| $L$ (mm) | 1.5 | 1 |
| On-chip loss (dB) | -4.8 | -11.0 |
| $\alpha V_\pi L$ (V·dB) | 0.0624 | 0.2464 |
| Switching time (ms) | < 1.5 | < 2 |

Along with smaller longitudinal size for ultracompact device design, the DC architecture is also smaller in lateral footprint. Assuming 5 μm of metal contact width and 4 μm of separation between metal contacts, the lateral sizes of the ultracompact DC and MZI switch will be 14 μm and 18 μm, respectively. With the demonstrated $V_\pi L$ and device lengths assumed in the previous paragraph, the DC configuration only occupies < 20% of device area compared with the MZI configuration. The small footprint and low loss characteristics make the DC architecture a leading candidate for ultralow-$V_\pi L$ photonic switch elements in cascaded switch arrays.

It is worth noting that the demonstrated device with 1.5 mm device length is highly wavelength dependent due to the small free spectral range (FSR) of ~4.5 nm. And the -4.8 dB loss is too high for cascaded switch arrays. However, the FSR is inversely proportional to the device length, and the propagation loss is proportional to the device length. The projected device with only 20-μm length is expected to possess a broadband FSR of ~337.5 nm and low-loss characteristics favorable for switch array applications.

Since the device is operated with 5 kHz square-wave ac voltages, the static power consumption to maintain optical signal level 0 and level 1, $P_i$ ($i = 0, 1$), can be calculated by

$$P_i = f \cdot C \cdot V_i^2, \quad (2)$$

where $f$ is the frequency of the ac voltage, $C$ is the capacitance of the slot capacitor, and $V_i$ is the corresponding voltage at the signal level ($i = 0, 1$). The measured $C$ at 5 kHz and 1 V is approximately 130 fF. Using $V_1$ we can estimate the upper limit of the total power consumption, including static and switching, to be 0.6 nW.

It is indicated in the power consumption analysis that a lower driving frequency is preferred to operate the device more power efficiently. Another consideration of the operation frequency is to maintain a fairly high frequency for stable optical signal levels. If the applied frequency is comparable to the NLC response speed, the NLC molecules will keep reorienting with the applied ac voltage and the optical signal levels will not be stable. In other words, there is a trade-off between the device speed and power consumption.

In this work we chose the widely used E7 as the NLC cladding to demonstrate the device. However, it is worth noting that the NLC material can be flexibly chosen based on different design considerations such as efficiency, switching speed, and power consumption. In general, NLC materials with high birefringence and low viscosity are preferable due to the resulting stronger and faster EO responses [25]. The dielectric anisotropy, elastic constants, and anchoring strength of NLC materials are also critical properties to achieve further improved efficiency and lower operating point [13], [14], [26].

Adding additives into NLC materials is another strategy to improve NLC material properties for EO applications. To incorporate NLC with additives into the proposed device, the additive unit size and slot width need to be considered. For instance, $C_{60}$ [27] and gold nanoparticles [28], as dopants, were reported to successfully improve the NLC EO performance. Although more studies are required to know if these materials provide the claimed improvement for the case within a silicon nano-slot, these methods are encouraging to further enhance the performance of the DC switch.

## VI. CONCLUSION

We have experimentally demonstrated a silicon photonic DC switch with 0.0195 V·mm modulation efficiency and 0.0624 V·dB loss-efficiency product. We have compared this DC switch performance with similar MZI switch device architectures. Without requiring a pair of combiner/splitter, additional routing waveguides, and strip-to-slot converters, we find the DC switch is superior to a corresponding MZI switch for efficient, compact optical switching due to a comparably smaller device footprint and lower optical loss. Improved DC performance can be expected as improved NLCs are incorporated into the switching devices.